\documentclass{cjaa}                   

\usepackage{graphicx}                   
\input{epsf.sty}                        
\input{psfig.sty}                       

\begin{document}

   \title{Microquasar models for 3EG J1828$+$0142 and 3EG J1735$-$1500}

   \volnopage{Vol.0 (200x) No.0, 000--000}      
   \setcounter{page}{1}          

   \author{V. Bosch-Ramon
      \inst{1}\mailto{vbosch@am.ub.es}
   \and J. M. Paredes
      \inst{1} 
   \and G. E. Romero
      \inst{2,3} 
   \and D. F. Torres
      \inst{4} 
      }
   \offprints{V. Bosch-Ramon}                   

   \institute{Departament d'Astronomia i Meteorologia, Universitat de Barcelona, 
              Av. Diagonal 647, E-08028 Barcelona, Spain\\
             \email{vbosch@am.ub.es}
        \and
             Instituto Argentino de Radioastronom\'{\i}a, C.C.5, (º894) Villa Elisa, 
	     Buenos Aires, Argentina
        \and
             Facultad de Ciencias Astron\'omicas y Geof\'{\i}sicas, UNLP, Paseo del Bosque, 
	     1900 La Plata, Argentina
	\and
	     Lawrence Livermore National Laboratory, 7000 East Avenue, L-413, 
	     Livermore, CA 94550 
          }

   \date{?; ?}

   \abstract{
Microquasars are promising candidates to emit high-energy gamma-rays.
Moreover, statistical studies show that variable EGRET sources at low galactic latitudes could be
associated with the inner spiral arms. The variable nature and the location in
the Galaxy of the high-mass microquasars, concentrated in the galactic plane
and within 55 degrees from the galactic center, give to these objects the
status of likely counterparts of the variable low-latitude EGRET sources. We
consider in this work the two most variable EGRET sources at low-latitudes: 
3EG~J1828$+$0142 and 3EG~J1735$-$1500, proposing a microquasar model to explain the
EGRET data in consistency with the observations at lower energies (from radio
frequencies to soft gamma-rays) within the EGRET error box.
   }

   \authorrunning{V. Bosch-Ramon et al.}           
   \titlerunning{Are 3EG J1828+0142 and 3EG J1735-1500 microquasars?}  

   \maketitle

\section{Introduction}           
\label{sect:intro}
Microquasars are X-ray binaries (XRB) that present relativistic jets (Mirabel \& Rodriguez
\cite{Mirabel99}). In the past decade, microquasars have also been proposed as high energy
sources, emitting into the the gamma-ray regime. For instance, Paredes et al.
(\cite{Paredes00}) proposed LS 5039, a high-mass microquasar, as the counterpart of the
source 3EG~J1824$-$1514. Otherwise, variable low-latitude EGRET sources seem to follow a
similar distribution to that presented by high-mass microquasars (Bosch-Ramon et al.
\cite{Bosch-Ramon04}), which are concentrated also in the galactic plane, not too far away
from the galactic center (55 degrees).  3EG~J1828$+$0142 and 3EG~J1735$-$1500 are the two
most variable EGRET sources in the galactic plane (Torres et al. \cite{Torres01}),
presenting quite steep spectra (photon indices of 2.76 and 3.24 respectively) and typical
luminosities of about 10$^{35}$~erg~s$^{-1}$, adopting a distance of 4 kpc. We have applied
a microquasar model to these particular cases for checking the proposal of association
between microquasars and the variable EGRET sources in the galactic plane.

\section{The Variable EGRET Sources in the Galactic Plane}
\label{sect:Thevar}
EGRET sources in the galactic plane are well-correlated with star forming regions (Romero et
al. \cite{Romero99}), and log N-log S studies suggest that they are more abundant toward the
inner spiral arms (Gehrels et al. \cite{Gehrels00}, Bhattacharya et al.
\cite{Bhattacharya03}). Among these sources, there is a subgroup of variable sources (Nolan
et al. \cite{Nolan03}) that has been proposed to be the EGRET counterparts of high-mass
microquasars (Kaufman Bernad\'o et al. \cite{Kaufman02}, Bosch-Ramon et al. 
\cite{Bosch-Ramon04}). High-mass microquasars are also
somehow correlated with star forming regions (Romero et al. \cite{Romero04}). We are
interested now in two particular cases: 3EG~J1828$+$0142 and 3EG~J1735$-$1500, exploring the
possibility that they can be high-mass microquasars.

\section{Gamma-ray Emission from Microquasars}
\label{sect:gamma}

\subsection{The Model}

We have developed a semi-analytical model based on an inhomogeneous leptonic jet formed by
relativistic particles. Those particles interact with the seed photon fields (synchrotron, star,
disk and corona photons) through inverse Compton (IC) effect (see Bosch-Ramon et al. 2004). We have
accounted for both the Thomson and the Klein-Nishina regimes of IC interaction (Blumenthal \& Gould
\cite{Blumenthal70}). The different functions that represent the electron energy distribution, the
electron energy and the magnetic field within the jet have been parametrized in order to simulate
their evolution along the jet (e.g. Ghisellini et al. \cite{Ghisellini85}, Punsly et al. 
\cite{Punsly00}).

\subsection{3EG J1828+0142 and 3EG J1735-1500}

3EG~J1828$+$0142 (Hartman et al. \cite{Hartman99}) is the second most variable low-latitude
non-transient gamma-ray source. Within the error box of this EGRET source, there are several
faint non-thermal radio sources (Punsly et al. 2000) and X-ray sources (observed in the
ROSAT All Sky Survey) with typical luminosities  of about 10$^{33}$~erg~s$^{-1}$, adopting
distances of 4~kpc. Finally, Comptel upper limits (Shu Zhang \cite{Zhang04}) are also known,
corresponding to luminosities of about 10$^{34}$-10$^{35}$~erg~s$^{-1}$ in the COMPTEL
energy range at the same distance. The error box of 3EG~J1735$-$1500 (Hartman et al.
\cite{Hartman99}), the most variable EGRET source, has been already explored by Combi et al.
(\cite{Combi03}), and there are two potential counterparts: a radio galaxy and a compact
radio source that presents hard spectrum and flux densities of about 0.3~Jy. High upper
limits at X-ray and COMPTEL energies are imposed from observational data to be 
(10$^{34}$-10$^{35}$~erg~s$^{-1}$). To model the Spectral Energy Distribution (SED) of a
microquasar that could be the origin of the EGRET emission, we account for the known
observational data at different wavelengths.  We must note that the observations at different
frequencies were not simultaneous. Regarding upper limits at radio and X-ray energies, they
have been taken as the fluxes of the brightest sources within the EGRET error boxes.

\begin{table}[]
  \caption[]{Common parameters for 3EG~J1828+0142 and 3EG~J1735-1500}
  \label{common}
  \begin{center}\begin{tabular}{clcl}
  \hline\noalign{\smallskip}
Parameter &  Value \\
  \hline\noalign{\smallskip}
Jet kinetic luminosity & 5$\times10^{35}$~erg~s$^{-1}$ \\
Stellar bolometric luminosity & $10^{38}$~erg~s$^{-1}$ \\
Distance from jet's apex to the compact object & $\sim 10^8$~cm \\  
Initial jet radius & $\sim 10^7$~cm \\
Orbital radius & 3$\times10^{12}$~cm \\ 
Viewing angle to jet's axis & $10^{\circ}$ \\  
Magnetic field & 300~G \\
  \noalign{\smallskip}\hline
  \end{tabular}\end{center}
\end{table}

\begin{table}[]
  \caption[]{Particular Parameters for 3EG~J1828+0142 and 3EG~J1735-1500}
  \label{part}
  \begin{center}\begin{tabular}{clcl}
  \hline\noalign{\smallskip}
Parameter & Adopted values for & Adopted values for \\
 & 3EG~J1828+0142 & 3EG~J1735-1500 \\
\hline\noalign{\smallskip}
Jet Lorentz factor & 1.5 & 3 \\
Maximum Lorentz factor for electrons in jet (jet frame) & 3$\times10^3$ & 2.5$\times10^3$ \\
Electron power-law index & 1.5 & 2 \\  
Total disk/corona luminosity & $10^{32}$~erg~s$^{-1}$ & $10^{33}$~erg~s$^{-1}$ \\
  \noalign{\smallskip}\hline
  \end{tabular}\end{center}
\end{table}

\section{Results}
\label{sect:res}

\begin{figure}
   \vspace{2mm}
   \begin{center}
   \hspace{3mm}\psfig{figure=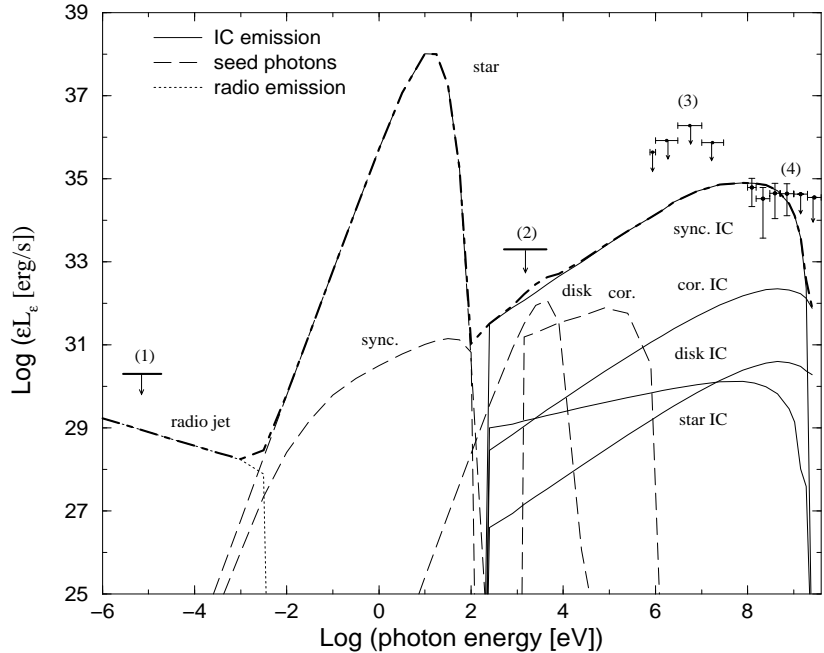,width=110mm,height=90mm,angle=0.0}
   \caption{SED for an unabsorbed broadband microquasar model of the source 3EG J1828+0142. 
The adopted parameter values are shown in Tables \ref{common} and \ref{part}. There are 
represented the upper limits at radio (1), X-ray (2) and COMPTEL (3) energies, as well 
as the EGRET spectrum (4).}
   \label{Fig:1828}
   \end{center}
\end{figure}

\begin{figure}
   \vspace{2mm}
   \begin{center}
   \hspace{3mm}\psfig{figure=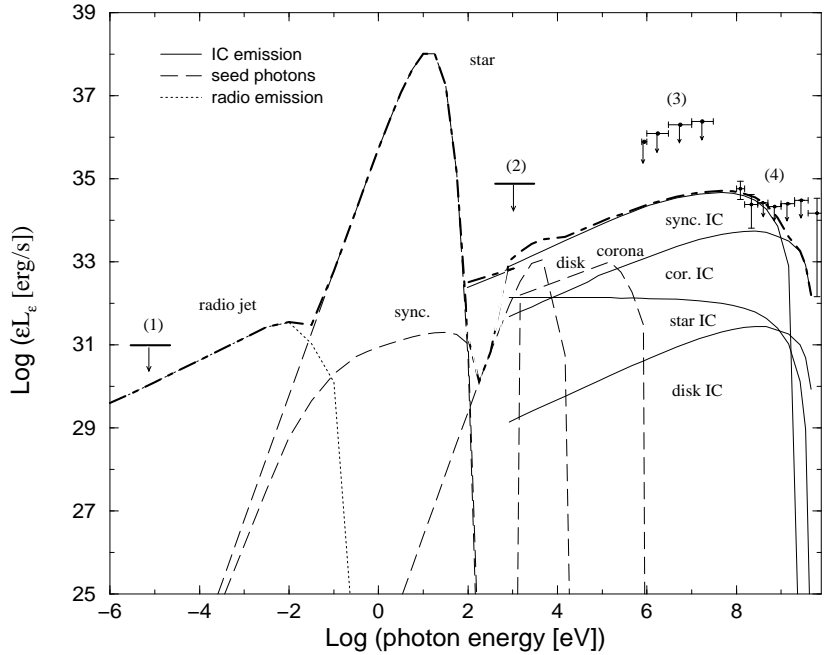,width=110mm,height=90mm,angle=0.0}
   \caption{The same as in Fig.~\ref{Fig:1828} but for the source 3EG J1735+1500.}
   \label{Fig:1735}
   \end{center}
\end{figure}

The Figs. \ref{Fig:1828} and \ref{Fig:1735} show the computed SED for 3EG~J1828$+$0142 and
3EG~J1735$-$1500 respectively. In Tables \ref{common} and \ref{part} the adopted parameters
for both systems are listed. These SED respect the observational constraints and reproduce
pretty well the EGRET spectrum. It is worth noting that the used EGRET spectrum for
comparison with our model is the averaged one for the four viewing periods. Nevertheless,
concerning variability, microquasars can naturally present the variable nature observed in
these two objects due to several factors: precession of the jet, orbital motion, etc...
(i.e. Kaufman Bernado et~al. \cite{Kaufman02}, Bosch-Ramon \& Paredes
\cite{Bosch-Ramon04b}).

Regarding the high-mass star emission of these systems, it could be dust-enshrouded, as it
has been suggested, for instance, for INTEGRAL obscured sources thought to be XRB and
microquasars (i.e. Walter et al. \cite{Walter03}, Combi et al. \cite{Combi04}). This
effect can be very important if the sources are located in the inner regions of the galactic
plane, being strong enough to obscure a high-mass star from IR wavelengths to soft X-rays.
Furthermore, emission in the far infrared can be too weak to be detected by the satellite
IRAS (i.e. Filliatre \& Chaty \cite{filli04}).

To compute the overall SED presented in the Figs. \ref{Fig:1828} and \ref{Fig:1735}, we have adopted a
strong magnetic field and a low maximum energy for the electrons (see Tables \ref{common} and
\ref{part}). For 3EG~J1828$+$0142, the synchrotron emission is dominant in the entire EGRET detection
range, which reached only several hundreds of MeV. Moreover, the jet is mildly relativistic. In the
other hand, 3EG~J1735$-$1500 was detected up to few GeV. For this second case, we have modelised the
emission up to several hundreds of MeV through synchrotron self-Compton scattering, and through
comptonization of coronal seed photons further up. For this second case, the jet Lorentz factor has
been taken higher than for the first one. For both sources, accounting for X-ray emission must be dim
since otherwise it would imply a clear counterpart at energies beyond 1~keV, corona and disk
components have been taken to be weak.

\section{Conclusions}
\label{sect:con}

The possibility of computing a radio-to gamma microquasar SED, at all in accord with 
observational constraints, suggests that microquasars may be the counterparts of
these two particular sources: 3EG~J1828$+$0142 and 3EG~J1735$-$1500. Other possible objects
cannot be discarded as possible counterparts (i.e. an isolated black-hole, see  Punsly et
al. \cite{Punsly00}; accreting neutron stars, see Romero et~al. \cite{Romero01b},; 
early-type binaries, see Benaglia \& Romero \cite{Benaglia03}; etc...). However,
microquasars appear to be attractive candidates due to the presence of a jet with
a relativistic leptonic population and strong seed photon fields provided by the stellar
companion, the accreting matter and the electrons themselves. In the near future,
instruments like AGILE and GLAST will help us to improve the location of unidentified EGRET
sources and further observations at lower energies will help us to better constrain the
models as well.


\begin{acknowledgements}

V.B-R. and J.M.P. acknowledge partial support by DGI of the Ministerio de Ciencia y
Tecnolog\'{\i}a (Spain) under grant AYA-2001-3092, as well as additional support from the
European Regional Development Fund (ERDF/FEDER). During this work, V.B-R has been supported
by the DGI of the Ministerio de Ciencia y Tecnolog\'{\i}a (Spain) under the fellowship
FP-2001-2699. G. E. R is supported by Fundaci\'on Antorchas and the Argentine Agencies
CONICET and ANPCyT. This research has benefited from a cooperation agreement supported by
Fundaci\'on Antorchas. The work of DFT was performed under the auspices of the US DOE (NNSA),
by UC's LLNL under contract No. W-7405-Eng-48. We thank an anonymous referee for useful
comments on the paper.

\end{acknowledgements}

\label{lastpage}

\end{document}